\documentclass[pra,print,showpacs,superscriptaddress,twocolumn]{revtex4}
\usepackage[dvips]{graphics,color}
\usepackage{amsfonts}
\usepackage{amsmath}
\usepackage{amssymb}
\usepackage{graphicx}
\usepackage{multirow}
\usepackage{float}

\begin{document}

\title{Two-axis spin squeezing in two cavities}
\author{Caifeng Li}
\affiliation{State Key Laboratory of Quantum Optics and Quantum Optics Devices, Institute
of Laser Spectroscopy, Shanxi University, Taiyuan 030006, China}
\affiliation{Institute of Theoretical Physics, Shanxi University, Taiyuan 030006, P. R.
China}
\author{Jingtao Fan}
\affiliation{State Key Laboratory of Quantum Optics and Quantum Optics Devices, Institute
of Laser Spectroscopy, Shanxi University, Taiyuan 030006, China}
\author{Lixuan Yu}
\thanks{yulixian@usx.edu.cn}
\affiliation{Department of Physics, Shaoxing University, Shaoxing 312000, China}
\affiliation{State Key Laboratory of Quantum Optics and Quantum Optics Devices, Institute
of Laser Spectroscopy, Shanxi University, Taiyuan 030006, China}
\author{Gang Chen}
\thanks{chengang971@163.com}
\affiliation{State Key Laboratory of Quantum Optics and Quantum Optics Devices, Institute
of Laser Spectroscopy, Shanxi University, Taiyuan 030006, China}
\author{Tian-Cai Zhang}
\affiliation{State Key Laboratory of Quantum Optics and Quantum Optics Devices, Institute
of Opto-Electronics, Shanxi University, Taiyuan 030006, China}
\author{Suotang Jia}
\affiliation{State Key Laboratory of Quantum Optics and Quantum Optics Devices, Institute
of Laser Spectroscopy, Shanxi University, Taiyuan 030006, China}

\begin{abstract}
Ultracold atoms in an ultrahigh-finesse optical cavity are a powerful
platform to produce spin squeezing since photon of cavity mode can induce
nonlinear spin-spin interaction and thus generate a one-axis twisting
Hamiltonian $H_{\text{OAT}}=qJ_{x}^{2}$, whose corresponding maximal
squeezing factor scales as $N^{-2/3}$, where $N$ is the atomic number. On the contrary, for the other
two-axis twisting Hamiltonian $H_{\text{TAT}}=q(J_{x}^{2}-J_{y}^{2})$, the
maximal squeezing factor scales as $N^{-1}$, approaching the Heisenberg
limit. In this paper, inspired by recent experiments of cavity-assisted
Raman transitions, we propose a scheme, in which an ensemble of ultracold
six-level atoms interacts with two quantized cavity fields and two pairs of
Raman lasers, to realize a tunable two-axis spin Hamiltonian $%
H=q(J_{x}^{2}+\chi J_{y}^{2})+\omega _{0}J_{z}$. For proper parameters, the
above one- and two- axis twisting Hamiltonians are recovered, and the
scaling of $N^{-1}$ of the maximal squeezing factor can occur naturally. On the other hand, in the two-axis twisting Hamiltonian, spin squeezing is usually reduced when increasing the effective atomic resonant frequency $\omega _{0}$. Surprisingly, we find that by combined with the dimensionless parameter $\chi(>-1)$, the effective atomic resonant frequency $\omega _{0}$ can enhance
spin squeezing largely. These results are benefit for achieving the
required spin squeezing in experiments.
\end{abstract}

\maketitle

\section{Introduction}

Squeezed spin states, which were firstly introduced by Kitagawa and Ueda
\cite{MK93}, are quantum correlated states with reduced fluctuations in one
of the collective spin components \cite{JM11,NPR13}. Such states have
attracted considerable interest because they not only play significant roles
in investigating many-body entanglement \cite{Dl00,AS01}, but also have
important applications in atom interferometers and high-precision atom
clocks \cite{SG99,KH10}. In general, there are two methods to produce spin
squeezing. One is based on a one-axis twisting Hamiltonian $H_{\text{OAT}%
}=qJ_{x}^{2}$, where $q$ is the nonlinear spin-spin interaction strength and
$J_{x}$ is the collective spin operator in the $x$ direction. When the
initial state is prepared as $\left\vert J_{z}=-j\right\rangle $ for $q>0$ ($%
\left\vert J_{z}=j\right\rangle $ for $q<0$), the maximal squeezing factor
for this one-axis twisting Hamiltonian scales as $N^{-2/3}$ \cite{MK93}, where $%
N$ is the total atomic number and $J=N/2$. On the contrary, for the other
two-axis twisting Hamiltonian $H_{\text{TAT}}=q(J_{x}^{2}-J_{y}^{2})$, the
maximal squeezing factor scales as $N^{-1}$ with the same initial state \cite%
{MK93}. Since the scaling of $N^{-1}$ approaches the Heisenberg limit,
implementing the two-axis twisting Hamiltonian in current experimental
setups is very important and necessary \cite{KM01,IB02,MZ03}. A proposing
scheme is to transform the one-axis twisting Hamiltonian into the two-axis
twisting Hamiltonian by applying pulse sequences or continuous driving in
the two-component Bose-Einstein condensates \cite{YCL11,CS13,JYZ14,WH14}.
However, the experimental realization of such two-axis twisting Hamiltonian
is still challenging.

Ultracold atoms in an ultrahigh-finesse optical cavity are also a powerful
platform to produce spin squeezing since photon of cavity mode can induce
nonlinear spin-spin interaction and thus generate the required one-axis
twisting Hamiltonian \cite{RND06,AEBN08,MHS10,IDL10,ZS11,ED13,ZC14,LY14}.
Recently, multi-mode cavities \cite{AM03} have attracted much attention both
experimentally and theoretically \cite{MM08,MM11,DJ13,AW13,DOK14}. On one
hand, these setups can be used to explore novel physics, such as the
spin-orbit-induced anomalous Hall effect \cite{JL09,JL10}, the
crystallization and frustration \cite{SG09,SG10}, the spin glass \cite%
{SG11,PS11,MB13,AA13}, and the gapless Nambu-Goldstone-type mode without
rotating-wave approximation \cite{JF14}. Moreover, two-mode field squeezing
\cite{RG06} and unconditional preparation of a two-mode squeezed state of
effective bosonic modes \cite{ASP06} have also been achieved by introducing
two cavities. In the present paper, inspired by recent experiments of
cavity-assisted Raman transitions \cite{KB10,MPB14}, we mainly realize a
generalized two-axis spin Hamiltonian by two cavities to enhance spin
squeezing largely.

When an ensemble of ultracold six-level atoms interacts with two quantized
cavity fields and two pairs of Raman lasers, we first realize a two-mode
Dicke model. In the dispersive regime, we obtain a generalized two-axis spin
Hamiltonian $H=q(J_{x}^{2}+\chi J_{y}^{2})+\omega _{0}J_{z}$, where $\chi $
is a dimensionless parameter and $\omega _{0}$ is an effective atomic
resonant frequency. This realized Hamiltonian has a distinct property that the
interaction strength $q$, the dimensionless parameter $\chi $, and the
effective atomic resonant frequency $\omega _{0}$ can be tuned
independently. For reasonable parameters, the one- and two-axis
twisting Hamiltonians are recovered. Numerical results reveal that for the standard
two-axis twisting Hamiltonian ($\chi =-1$ and $\omega _{0}=0$), the
corresponding maximal squeezing factor scales as $N^{-1}$, as expected. On the other hand, in the two-axis twisting Hamiltonian $H_{\text{TAT}}$, spin squeezing is usually reduced when increasing the effective atomic resonant frequency $\omega _{0}$. Surprisingly, we find that by combined with the dimensionless parameter $\chi(>-1)$, the effective atomic resonant frequency $\omega _{0}$ can enhance
spin squeezing largely. These results are benefit for achieving the required
spin squeezing in experiments.

This paper is organized as follows. Section II is devoted to realizing the
generalized two-axis spin Hamiltonian with independently-tunable parameters.
Section III is devoted to introducing the spin squeezing factor. Section IV
is devoted to numerically investigating the maximal squeezing factor since
for the two-axis spin Hamiltonian, analytical results are very hard to be
obtained \cite{JM11}. The parts of Discussions and Conclusions are given in
sections V and VI.

\section{Model and Hamiltonian}

\begin{figure}[tp]
\includegraphics[width=7cm]{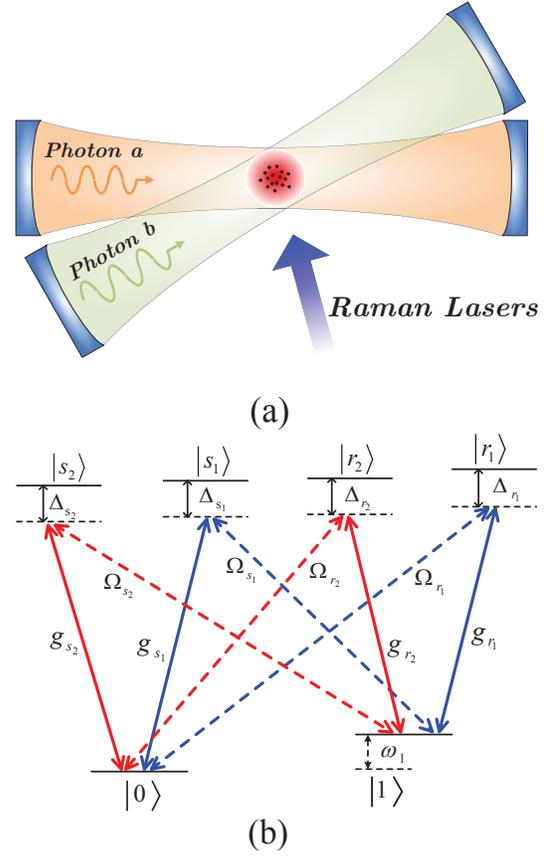}\newline
\caption{(Color online). (a) Proposed experimentally-feasible setup that an
ensemble of ultracold six-level atoms interacts with two quantized cavity
fields and two pairs of Raman lasers. (b) Energy-level structure and their
transitions induced by two photon modes and two pairs of Raman lasers.}
\label{fig1}
\end{figure}

\subsection{Proposed experimental setup}

Motivated by recent experiments of cavity-assisted Raman transitions \cite%
{KB10,MPB14}, here we propose a scheme, in which an ensemble of ultracold
six-level atoms interacts with two quantized cavity fields and two pairs of
Raman lasers [see Fig.~\ref{fig1}(a)], to realize a generalized two-axis
spin Hamiltonian with independently-tunable parameters. As shown in Fig.~\ref%
{fig1}(b), six levels consist of two stable ground states ($\left\vert
0\right\rangle $ and $\left\vert 1\right\rangle $) and four excited states ($%
\left\vert r_{1}\right\rangle $, $\left\vert r_{1}\right\rangle $, $%
\left\vert s_{1}\right\rangle $, and $\left\vert s_{2}\right\rangle $). Two
independent photon modes, whose creation and annihilation operators\ are $%
a^{\dag }$ $(b^{\dag })$ and $a$ $(b)$, mediate the $\left\vert
0\right\rangle \longleftrightarrow \left\vert s_{1}\right\rangle $ and $%
\left\vert 1\right\rangle \longleftrightarrow \left\vert r_{1}\right\rangle $
($\left\vert 0\right\rangle \longleftrightarrow \left\vert
s_{2}\right\rangle $ and $\left\vert 1\right\rangle \longleftrightarrow
\left\vert r_{2}\right\rangle $) transitions (red and blue solid lines) with
atom-photon coupling strengths $g_{s_{1}}$ $(g_{s_{2}})$ and $g_{r_{1}}$ $%
(g_{r_{2}})$, respectively. Whereas two pairs of Raman lasers govern the
other transitions \{$\left\vert 0\right\rangle \longleftrightarrow
\left\vert r_{1}\right\rangle $, $\left\vert 1\right\rangle \leftrightarrow
\left\vert s_{1}\right\rangle $\} and \{$\left\vert 0\right\rangle
\longleftrightarrow \left\vert r_{2}\right\rangle $, $\left\vert
1\right\rangle \leftrightarrow \left\vert s_{2}\right\rangle $\} (red and
blue dashed lines) with Rabi frequencies \{$\Omega _{r_{1}}$, $\Omega
_{s_{1}}$\} and \{$\Omega _{r2}$, $\Omega _{s_{2}}$\}, respectively. $\Delta
_{r_{1,2}}$ and $\Delta _{s_{1,2}}$ are the detunings from the excited
states.

\subsection{Total Hamiltonian}

The total Hamiltonian illustrated in Fig.~\ref{fig1} can be written as
\begin{equation}
H_{\text{T}}=H_{\text{C}}+H_{\text{A}}+H_{\text{I}},  \label{Ha}
\end{equation}%
where
\begin{equation}
H_{\text{C}}=\omega _{a}a^{\dag }a+\omega _{b}b^{\dagger }b,  \label{Hcav}
\end{equation}%
\begin{align}
H_{\text{A}}& =\sum_{j=1}^{N}\left\{ \omega _{r_{1}}\left\vert
r_{1}\right\rangle _{j}\left\langle r_{1}\right\vert _{j}+\omega
_{r_{2}}\left\vert r_{2}\right\rangle _{j}\left\langle r_{2}\right\vert
_{j}+\right.  \label{Hat} \\
& +\omega _{s_{1}}\left\vert s_{1}\right\rangle _{j}\left\langle
s_{1}\right\vert _{j}+\omega _{s_{2}}\left\vert s_{2}\right\rangle
_{j}\left\langle s_{2}\right\vert _{j}+\omega _{1}\left\vert 1\right\rangle
_{j}\left\langle 1\right\vert _{j}  \notag \\
& +\frac{1}{2}\left[ \Omega _{s_{1}}\left\vert s_{1}\right\rangle
_{j}\left\langle 1\right\vert _{j}e^{-i(\omega _{1s_{1}}t-\varphi
_{s_{1}})}\right.  \notag \\
& +\Omega _{s_{2}}\left\vert s_{2}\right\rangle _{j}\left\langle
1\right\vert _{j}e^{-i(\omega _{1s_{2}}t-\varphi _{s_{2}})}  \notag \\
& +\Omega _{r_{1}}\left\vert r_{1}\right\rangle _{j}\left\langle
0\right\vert _{j}e^{-i(\omega _{0r_{1}}t-\varphi _{r_{1}})}  \notag \\
& +\left. \left. \Omega _{r_{2}}\left\vert r_{2}\right\rangle
_{j}\left\langle 0\right\vert _{j}e^{-i(\omega _{0r_{2}}t-\varphi _{r_{2}})}+%
\text{H.c.}\right] \right\} ,  \notag
\end{align}%
\begin{eqnarray}
H_{\text{I}} &=&\sum_{j=1}^{N}\left[ g_{s_{1}}\left\vert s_{1}\right\rangle
_{j}\left\langle 0\right\vert _{j}a+g_{r_{1}}\left\vert r_{1}\right\rangle
_{j}\left\langle 1\right\vert _{j}a\right.  \label{Hint} \\
&&\left. +g_{s_{2}}\left\vert s_{2}\right\rangle _{j}\left\langle
0\right\vert _{j}b+g_{r_{2}}\left\vert r_{2}\right\rangle _{j}\left\langle
1\right\vert _{j}b+\text{H.c.}\right] .  \notag
\end{eqnarray}%
In the Hamiltonians~(\ref{Hcav})-(\ref{Hint}), $\omega _{r_{1,2}}$, $\omega
_{s_{1,2}}$, and $\omega _{1}$ are the atomic frequencies, $\omega
_{1s_{1,2}}$ ($\varphi _{s_{1,2}}$) and $\omega _{1r_{1,2}}$ ($\varphi
_{r_{1,2}}$) are the frequencies (phases) of Raman lasers, respectively, and
H.c. denotes the Hermitian conjugate.

\subsection{Two-mode Dicke model}

By means of the Hamiltonian~(\ref{Ha}), we first realize a two-mode Dicke
model with independently-tunable parameters. In the interaction picture with
respect to the free Hamiltonian
\begin{eqnarray}
H_{0} &=&\widetilde{\omega _{a}}a^{\dag }a+\widetilde{\omega _{b}}b^{\dagger
}b+\sum_{j=1}^{N}\left( \widetilde{\omega _{s_{1}}}\left\vert
s_{1}\right\rangle _{j}\left\langle s_{1}\right\vert _{j}\right.
\label{INTHA} \\
&&+\widetilde{\omega _{r_{1}}}\left\vert r_{1}\right\rangle _{j}\left\langle
r_{1}\right\vert _{j}+\widetilde{\omega _{s_{2}}}\left\vert
s_{2}\right\rangle _{j}\left\langle s_{2}\right\vert _{j}  \notag \\
&&\left. +\widetilde{\omega _{r_{2}}}\left\vert r_{2}\right\rangle
_{j}\left\langle r_{2}\right\vert _{j}+\widetilde{\omega _{1}}\left\vert
1\right\rangle _{j}\left\langle 1\right\vert _{j}\right) ,  \notag
\end{eqnarray}%
where $\widetilde{\omega _{a}}=\widetilde{\omega _{s_{1}}}=(\omega
_{0r_{1}}+\omega _{1s_{1}})/2$, $\widetilde{\omega _{b}}=\widetilde{\omega
_{s_{2}}}=(\omega _{0r_{2}}+\omega _{1s_{2}})/2$, $\widetilde{\omega _{1}}%
=(\omega _{0r_{1}}-\omega _{1s_{1}})/2=(\omega _{0r_{2}}-\omega _{1s_{2}})/2$%
, $\widetilde{\omega _{r_{1}}}=\widetilde{\omega _{1}}+\widetilde{\omega _{a}%
}$, and $\widetilde{\omega _{r_{2}}}=\widetilde{\omega _{1}}+\widetilde{%
\omega _{b}}$, the Hamiltonian~(\ref{Ha}) can be rewritten as
\begin{eqnarray}
\widetilde{H} &=&\Delta _{a}a^{\dag }a+\Delta _{b}b^{\dagger
}b+\sum_{j=1}^{N}\left\{ \Delta _{r_{1}}\left\vert r_{1}\right\rangle
_{j}\left\langle r_{1}\right\vert _{j}\right.  \label{INTH} \\
&&+\Delta _{r_{2}}\left\vert r_{2}\right\rangle _{j}\left\langle
r_{2}\right\vert _{j}+\Delta _{s_{1}}\left\vert s_{1}\right\rangle
_{j}\left\langle s_{1}\right\vert _{j}+\Delta _{s_{2}}\left\vert
s_{2}\right\rangle _{j}\left\langle s_{2}\right\vert _{j}  \notag \\
&&+\Delta _{1}\left\vert 1\right\rangle _{j}\left\langle 1\right\vert _{j}+%
\frac{1}{2}\left[ \Omega _{s_{1}}\left\vert s_{1}\right\rangle
_{j}\left\langle 1\right\vert _{j}e^{i\varphi _{s_{1}}}\right.  \notag \\
&&+\Omega _{s_{2}}\left\vert s_{2}\right\rangle _{j}\left\langle
1\right\vert _{j}e^{i\varphi _{s_{2}}}+\Omega _{r_{1}}\left\vert
r_{1}\right\rangle _{j}\left\langle 0\right\vert _{j}e^{i\varphi _{r_{1}}}
\notag \\
&&\left. +\Omega _{r_{2}}\left\vert r_{2}\right\rangle _{j}\left\langle
0\right\vert _{j}e^{i\varphi _{r_{2}}}+\text{H.c.}\right] +\left[
g_{s_{1}}\left\vert s_{1}\right\rangle _{j}\left\langle 0\right\vert
_{j}a\right.  \notag \\
&&+g_{r_{1}}\left\vert r_{1}\right\rangle _{j}\left\langle 1\right\vert
_{j}a+g_{s_{2}}\left\vert s_{2}\right\rangle _{j}\left\langle 0\right\vert
_{j}b  \notag \\
&&\left. \left. +g_{r_{2}}\left\vert r_{2}\right\rangle _{j}\left\langle
1\right\vert _{j}b+\text{H.c.}\right] \right\} ,  \notag
\end{eqnarray}%
where $\Delta _{a}=\omega _{a}-\widetilde{\omega _{a}}$, $\Delta _{b}=\omega
_{b}-\widetilde{\omega _{b}}$, $\Delta _{s_{1}}=\omega _{s_{1}}-\widetilde{%
\omega _{s_{1}}}$, $\Delta _{s_{2}}=\omega _{s_{2}}-\widetilde{\omega
_{s_{2}}}$, and $\Delta _{1}=\omega _{1}-\widetilde{\omega _{1}}$.

In the large-detuning limit, i.e., $|\Delta _{r_{1,2},s_{1,2}}|\gg \{\Omega
_{r_{1,2}}$, $\Omega _{s_{1,2}}$, $g_{r_{1,2}}$, $g_{s_{1,2}}\}$, all
excited states can be eliminated adiabatically \cite{FD07}, and an effective
Hamiltonian is obtained by
\begin{eqnarray}
\widetilde{H} &=&\omega _{A}a^{\dag }a+\omega _{B}b^{\dag }b+\omega
_{0}J_{z}+\eta a^{\dag }aJ_{z}  \label{EFF} \\
&&+\left[ \left( \lambda _{r_{1}}ae^{-i\varphi _{r_{1}}}+\lambda
_{r_{2}}be^{-i\varphi _{r_{2}}}\right) J_{-}\right.  \notag \\
&&\left. +\left( \lambda _{s_{1}}ae^{-i\varphi _{s_{1}}}+\lambda
_{s_{2}}be^{-i\varphi _{s_{2}}}\right) J_{+}+\text{H.c.}\right] ,  \notag
\end{eqnarray}%
where $\eta =g_{r_{1}}^{2}/\Delta _{r_{1}}+g_{r_{2}}^{2}/\Delta
_{r_{2}}-g_{s_{1}}^{2}/\Delta _{s_{1}}-g_{s_{2}}^{2}/\Delta _{s_{2}}$, $%
J_{+}=\sum_{j=1}^{j=N}\left\vert 1\right\rangle _{j}\left\langle
0\right\vert _{j}$, $J_{-}=\sum_{j=1}^{j=N}\left\vert 0\right\rangle
_{j}\left\langle 1\right\vert _{j}$, and $J_{z}=\sum_{j=1}^{j=N}(\left\vert
1\right\rangle _{j}\left\langle 0\right\vert _{j}-\left\vert 0\right\rangle
_{j}\left\langle 0\right\vert _{j})/2$ are the collective spin operators,
\begin{equation}
\omega _{0}=\Delta _{1}+\frac{1}{4}\left( \frac{\Omega _{s_{1}}^{2}}{\Delta
_{s_{1}}}+\frac{\Omega _{s_{2}}^{2}}{\Delta _{s_{2}}}-\frac{\Omega
_{r_{1}}^{2}}{\Delta _{r_{1}}}-\frac{\Omega _{r_{2}}^{2}}{\Delta _{r_{2}}}%
\right) ,  \label{ARF}
\end{equation}%
\begin{equation}
\omega _{A}=\Delta _{a}+\frac{1}{2}\left( \frac{Ng_{r_{1}}^{2}}{\Delta _{r1}}%
+\frac{Ng_{s_{1}}^{2}}{\Delta _{s1}}\right) ,  \label{PRF1}
\end{equation}%
\begin{equation}
\omega _{B}=\Delta _{b}+\frac{1}{2}\left( \frac{Ng_{r_{2}}^{2}}{\Delta
_{r_{2}}}+\frac{Ng_{s_{2}}^{2}}{\Delta _{s_{2}}}\right) ,  \label{PRF2}
\end{equation}%
are the effective atomic resonant frequency and the effective frequencies of
two photon modes $a$ and $b$, respectively, and
\begin{equation}
\lambda _{r_{i}}=\frac{1}{2}\frac{g_{r_{i}}\Omega _{r_{i}}}{\Delta _{r_{i}}}%
,\lambda _{s_{i}}=\frac{1}{2}\frac{g_{s_{i}}\Omega _{s_{i}}}{\Delta _{s_{i}}}%
,(i=1,2)\text{ }  \label{IP}
\end{equation}%
are the effective atom-photon coupling strengths.

\begin{figure}[tp]
\includegraphics[width=8cm]{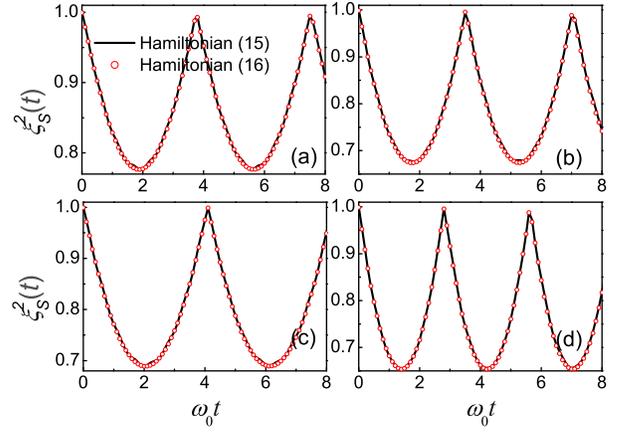}\newline
\caption{(Color online). Numerical plots of the time-dependent spin
squeezing factors $\protect\xi _{\text{S}}^{2}(t)$ of both the Hamiltonians (%
\protect\ref{HREF}) and (\protect\ref{TS}), with the same initial states $%
\left\vert J_{z}=-j\right\rangle $. In (a) and (b), $\protect\omega _{A}/%
\protect\omega _{0}=200$, $\protect\omega _{B}/\protect\omega _{0}=\pm 200$
[\textquotedblleft $+"$ for (a) and \textquotedblleft $-"$ for (b)], $%
\protect\lambda _{1}/\protect\omega _{0}=2$, $\protect\lambda _{2}/\protect%
\omega _{0}=1$, and $N=15$. In (c) and (d), $\protect\omega _{A}/\protect%
\omega _{0}=200$, $\protect\omega _{B}/\protect\omega _{0}=\pm 200$
[\textquotedblleft $+"$ for (c) and \textquotedblleft $-"$ for (d)], $%
\protect\lambda _{1}/\protect\omega _{0}=1$, $\protect\lambda _{2}/\protect%
\omega _{0}=2$, and $N=20$.}
\label{fig2}
\end{figure}

When choosing $g_{r_{i}}^{2}/\Delta _{r_{i}}=g_{s_{i}}^{2}/\Delta _{s_{i}}$,
$\Omega _{r_{i}}g_{r_{i}}/\Delta _{r_{i}}=\Omega _{s_{i}}g_{s_{i}}/\Delta
_{s_{i}}$, and $\varphi _{s_{i}}=-\varphi _{r_{i}}=\varphi _{i}$, the
Hamiltonian~(\ref{EFF}) turns into
\begin{align}
\widetilde{H}& =\omega _{A}a^{\dag }a+\omega _{B}b^{\dag }b+\omega
_{0}J_{z}+\lambda _{1}(a^{\dagger }+a)  \label{EFFH} \\
& \times \left( J_{+}e^{-i\varphi _{1}}+J_{-}e^{i\varphi _{1}}\right)
+\lambda _{2}(J_{+}e^{-i\varphi _{2}}+J_{-}e^{i\varphi _{2}})(b^{\dagger
}+b),  \notag
\end{align}%
where the effective atom-photon coupling strengths become
\begin{equation}
\lambda _{1}=\frac{1}{2}\frac{\Omega _{s_{1}}g_{s_{1}}}{\Delta _{s_{1}}}=%
\frac{1}{2}\frac{\Omega _{r_{1}}g_{r_{1}}}{\Delta _{r_{1}}},  \label{R1}
\end{equation}%
\begin{equation}
\lambda _{2}=\frac{1}{2}\frac{\Omega _{s_{2}}g_{s_{2}}}{\Delta _{s_{2}}}=%
\frac{1}{2}\frac{\Omega _{r_{2}}g_{r_{2}}}{\Delta _{r_{2}}}.  \label{R2}
\end{equation}%
If setting $\varphi _{1}=0$ and $\varphi _{2}=-\pi /2$, the Hamiltonian (\ref%
{EFFH}) becomes%
\begin{align}
\widetilde{H}& =\omega _{A}a^{\dag }a+\omega _{B}b^{\dag }b+\omega _{0}J_{z}
\label{HREF} \\
& +\lambda _{1}J_{x}(a^{\dagger }+a)+\lambda _{2}J_{y}(b^{\dagger }+b).
\notag
\end{align}%
The Hamiltonian~(\ref{HREF}) is our required two-mode Dicke model, based on
recent experiments of cavity-assisted Raman transitions \cite{KB10,MPB14}.
In contrast to the convectional two-mode Dicke model achieved in the
two-level atoms, the Hamiltonian~(\ref{HREF}) has a distinct property that
all parameters can be tuned independently. For example, the effective cavity
frequencies $\omega _{A}$ and $\omega _{B}$ depend on the detunings $\Delta
_{a}$ and $\Delta _{b}$, respectively; see Eqs.~(\ref{PRF1}) and~(\ref{PRF2}%
). Thus, they can range from the positive to the negative. The choice of the
different cavity frequencies $\omega _{A}$ and $\omega _{B}$ help us to
create a tunable two-axis spin Hamiltonian, as will be shown below. The
effective atomic resonant frequency $\omega _{0}$ can also be controlled by
the detuning $\Delta _{1}$; see Eq.~(\ref{ARF}). In addition, the effective
atom-photon coupling strengths $\lambda _{1}$ and $\lambda _{2}$ can be
driven by the Rabi frequencies of Raman lasers; see Eqs.~(\ref{R1}) and~(\ref%
{R2}).

\subsection{Generalized two-axis spin Hamiltonian}

In the following, we mainly consider the dispersive regime, i.e., $%
\{\left\vert \omega _{A}\right\vert ,\left\vert \omega _{B}\right\vert \}\gg
\{\lambda _{1},\lambda _{2}\}$. In such case, the photons are virtually
excited, and we can use the Heisenberg equations of motion \cite{JLE10}, $%
\dot{a}=-i(\lambda _{1}J_{x}+\omega _{A}a)=0$ and $\dot{b}=-i(\lambda
_{2}J_{y}+\omega _{B}b)=0$, to obtain $a=-\lambda _{1}J_{x}$\-$/\omega _{A}$
and $b=-\lambda _{2}J$\-$_{y}/\omega _{B}$. As a result, the Hamiltonian (%
\ref{HREF}) becomes
\begin{equation}
H=q(J_{x}^{2}+\chi J_{y}^{2})+\omega _{0}J_{z},  \label{TS}
\end{equation}%
where
\begin{equation}
q=-\frac{\lambda _{1}^{2}}{\omega _{A}},  \label{PA1}
\end{equation}%
\begin{equation}
\chi =\frac{\omega _{A}\lambda _{2}^{2}}{\omega _{B}\lambda _{1}^{2}}.
\label{PA2}
\end{equation}%
In the Hamiltonian~(\ref{TS}), the parameter $q$ determines the nonlinear
spin-spin interaction $J_{x}^{2}$ induced by the virtual photon, and the
dimensionless parameter $\chi $ reflects the ratio between the different
nonlinear spin-spin interactions $J_{x}^{2}$ and $J_{y}^{2}$. Due to
existence of these nonlinear spin-spin interactions with the
dimensionless parameter $\chi $ and the effective atomic resonant frequency $\omega _{0}$, here we call the
Hamiltonian~(\ref{TS}) as a generalized two-axis spin Hamiltonian. In
addition, equations (\ref{ARF}), (\ref{PA1}), and (\ref{PA2}) show clearly
that all the parameters, including the interaction strength $q$, the dimensionless parameter $\chi $, and the effective atomic resonant frequency $\omega _{0}$, can also be tuned
independently in experiments.

When chosen reasonable parameters, the generalized two-axis spin Hamiltonian
(\ref{TS}) can reduce to some well-studied Hamiltonians. For example, when $%
\chi >0$, the Hamiltonian~(\ref{TS}) is a standard Lipkin-Meshkov-Glick
model \cite{LMG1,LMG2,LMG3}. When $\omega _{0}=0$, the Hamiltonian~(\ref{TS}%
) reduces to a generalized two-axis twisting Hamiltonian $H_{\text{GTAT}%
}=q(J_{x}^{2}+\chi J_{y}^{2})$. If further setting $\chi =-1$, a standard
two-axis twisting Hamiltonian $H_{\text{TAT}}=q(J_{x}^{2}-J_{y}^{2})$ is
derived. Finally, when $\chi =0$, the Hamiltonian~(\ref{TS}) turns into a
generalized one-axis twisting Hamiltonian $H_{\text{GOAT}}=qJ_{x}^{2}+\omega
_{0}J_{z}$ \cite{CKL01}, which reduces to the standard one-axis twisting
Hamiltonian $H_{\text{OAT}}=qJ_{x}^{2}$ for $\omega _{0}=0$. These results
implies that the Hamiltonian~(\ref{TS}) has an important application in
achieving the required spin squeezing.

\begin{figure}[tp]
\includegraphics[width=8cm]{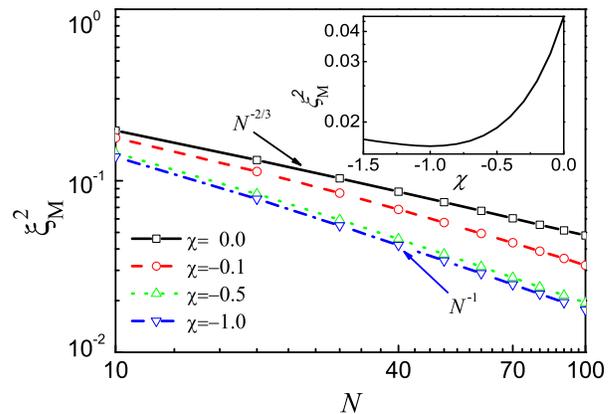}\newline
\caption{(Color online). Numerical plots of the maximal squeezing factors $%
\protect\xi _{\text{M}}^{2}$ as a function of the atomic number $N$ for the
different dimensionless parameters $\protect\chi $. Insert: Numerical plot
of the maximal squeezing factor $\protect\xi _{\text{M}}^{2}$ as a function
of the dimensionless parameter $\protect\chi $, when the atomic number is
chosen as $N=100$. In both subfigures, the initial states are chosen as $%
\left\vert J_{z}=-j\right\rangle $.}
\label{fig3}
\end{figure}

Notice that when $\omega _{A}<0$, the effective spin-spin interaction
strength $q>0$. Thus, we can use the initial state $\left\vert
J_{z}=-j\right\rangle $ to discuss spin squeezing of the generalized
two-axis spin Hamiltonian (\ref{TS}). This initial state $\left\vert
J_{z}=-j\right\rangle $ can be easily prepared in experiments. In addition,
by combined with the effective atomic resonant frequency $\omega _{0}$, the
dimensionless parameter $\chi $ plays an important role in spin squeezing,
as will be shown.

\section{Spin squeezing factor}

In order to investigate spin squeezing, it is very necessary to consider the
following time-dependent squeezing factor \cite{MK93}:
\begin{equation}
\xi _{\text{S}}^{2}(t)=\frac{4}{N}\min \left[ \Delta J_{\vec{n}_{\perp
}}^{2}(t)\right] ,  \label{SSS}
\end{equation}%
where $\vec{n}_{\bot }$ refers to an axis, which is perpendicular to the
mean-spin direction $\vec{n}_{0}=\vec{J}/\left\vert J\right\vert $ with $%
\left\vert J\right\vert =\sqrt{\left\langle J_{x}\right\rangle
^{2}+\left\langle J_{y}\right\rangle ^{2}+\left\langle J_{z}\right\rangle
^{2}}$, and $\Delta A^{2}=\left\langle A^{2}\right\rangle -\left\langle
A\right\rangle ^{2}$ is the standard deviation. If $\left\vert \xi _{\text{S}%
}^{2}(t)\right\vert <1$, the spin state is squeezed, and vice versa.

In the spherical coordinates, $\vec{n}_{0}=(\sin \theta \cos \varphi ,\sin \theta
\sin \varphi ,\cos \theta )$, where $\theta =\arccos (\left\langle
J_{z}\right\rangle /\left\vert J\right\vert )$, and $\varphi =\arccos \left(
\left\langle J_{x}\right\rangle /\left\vert J\right\vert \sin \theta \right)
$ for $\left\langle J_{y}\right\rangle >0$ or $\varphi =2\pi -\arccos \left(
\left\langle J_{x}\right\rangle /\left\vert J\right\vert \sin \theta \right)
$ for $\left\langle J_{y}\right\rangle \leq 0$. Two orthogonal bases are
given by $\vec{n}_{1}=(-\sin \varphi ,\cos \varphi ,0)$ and $\vec{n}%
_{2}=\left( -\cos \theta \cos \varphi ,\cos \theta \sin \varphi ,-\sin
\theta \right) $. Thus, $J_{\vec{n}_{1,2}}=\vec{J}\cdot \vec{n}_{1,2}$, $J_{%
\vec{n}_{\perp }}=J_{\vec{n}_{1}}\cos \phi +J_{\vec{n}_{2}}\sin \phi $, and $%
\min (\Delta J_{\vec{n}_{\perp }}^{2})$ could be achieved when $\phi $
varies from $0$ to $2\pi $ in the plane that is perpendicular to the
mean-spin direction $\vec{n}_{0}$. It should be noticed that in experiments, the maximal squeezing factor
$\xi _{\text{M}}^{2}$ is usually measured \cite{JM11,NPR13}. Thus, in the following discussions, we mainly focus on this physical quantity.

Before proceeding, we check the validity of the Hamiltonian~(\ref{TS}), when
the initial state is chosen as $\left\vert J_{z}=-j\right\rangle $. For the
generalized two-axis spin Hamiltonian, it is very hard to obtain analytical
result of the spin squeezing factor $\xi _{\text{S}}^{2}(t)$ \cite{JM11}. In
Fig.~\ref{fig2}, we numerically plot the corresponding spin squeezing
factors $\xi _{\text{S}}^{2}(t)$ of the Hamiltionians (\ref{HREF}) and (\ref%
{TS}). It can be seen clearly that the results of the Hamiltonian~(\ref{TS})
are almost identical to those of the Hamiltonian~(\ref{HREF}). Therefore, we will apply the Hamiltonian~(\ref{TS}) to
discuss the experimentally-measurable maximal squeezing factor $\xi _{\text{M}}^{2}$ in the rest of this paper.

\begin{figure}[tp]
\includegraphics[width=8cm]{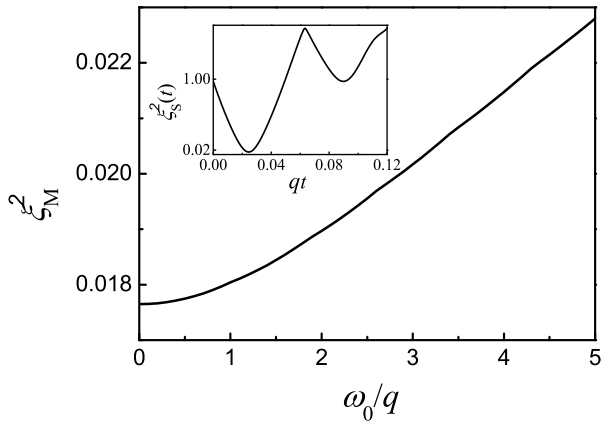}\newline
\caption{(Color online). Numerical plot of the maximal squeezing factor $%
\protect\xi _{\text{M}}^{2}$ as a function of the effective atomic resonant
frequency $\protect\omega _{0}$. Insert: Numerical plot of the
time-dependent squeezing factor $\protect\xi _{\text{S}}^{2}(t)$. In both
subfigures, the initial states, the dimensionless parameter, and the atomic
number are chosen as $\left\vert J_{z}=-j\right\rangle $, $\protect\chi =-1$%
, and $N=100$, respectively.}
\label{fig4}
\end{figure}

\section{Maximal squeezing factor}

We first address a simple case without the effective atomic resonant
frequency ($\omega _{0}=0$), in which the generalized two-axis spin
Hamiltonian~(\ref{TS}) reduces to the generalized two-axis twisting
Hamiltonian $H_{\text{GTAT}}=q(J_{x}^{2}+\chi J_{y}^{2})$. In Fig.~\ref{fig3}%
, we numerically plot the maximal squeezing factor $\xi _{\text{M}}^{2}$ of
the Hamiltonian $H_{\text{GTAT}}$ as a function of the atomic number $N$ for
the different dimensionless parameters $\chi $, when the initial state is
chosen as $\left\vert J_{z}=-j\right\rangle $. This figure shows clearly
that when $\chi =0$, the generalized two-axis twisting Hamiltonian $H_{\text{%
GTAT}}$ becomes the standard one-axis twisting Hamiltonian $H_{\text{OAT}%
}=qJ_{x}^{2}$, whose maximal squeezing factor $\xi _{\text{M}}^{2}$ scales
as $N^{-2/3}$ \cite{MK93}. When increasing the dimensionless parameter $\chi
$, the maximal squeezing factor $\xi _{\text{M}}^{2}$ decreases, i.e., spin
squeezing is enhanced. In particular, when $\chi =-1$, the Hamiltonian $H_{%
\text{GTAT}}$ turns into the standard two-axis twisting Hamiltonian $H_{%
\text{TAT}}=q(J_{x}^{2}-J_{y}^{2})$, whose maximal squeezing factor $\xi _{%
\text{M}}^{2}$ scales as $N^{-1}$ \cite{MK93}, as expected. In addition, the
maximal squeezing factor $\xi _{\text{M}}^{2}$ as a function of the
dimensionless parameter $\chi $ is also plotted in the insert part of Fig.~%
\ref{fig3}. This figure shows that $\chi =-1$ is an optimal point to achieve
the maximal squeezing factor $\xi _{\text{M}}^{2}$ of the generalized
two-axis twisting Hamiltonian $H_{\text{GTAT}}$.

In real experiments, the effective atomic resonant frequency $\omega _{0}$
always exists. In Fig.~\ref{fig4}, we numerically plot the maximal squeezing
factor $\xi _{\text{M}}^{2}$ of the Hamiltonian $H=q(J_{x}^{2}-J_{y}^{2})+%
\omega _{0}J_{z}$ as a function of the effective atomic resonant frequency $%
\omega _{0}$, when the the initial state is chosen as $\left\vert
J_{z}=-j\right\rangle $. It can be seen from this figure that with the
increasing of the effective atomic resonant frequency $\omega _{0}$ in the
standard two-axis twisting Hamiltonian $H_{\text{TAT}}$, the maximal
squeezing factor $\xi _{\text{M}}^{2}$ increases, i.e., spin squeezing is
reduced.

\begin{figure}[tp]
\includegraphics[width=7.5cm]{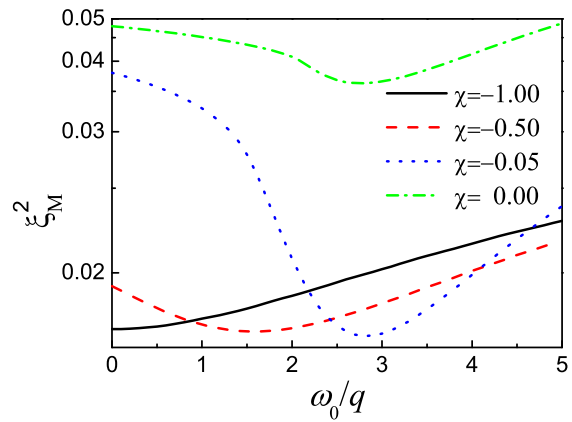}\newline
\caption{(Color online). Numerical plots of the maximal squeezing factors $%
\protect\xi _{\text{M}}^{2}$ as a function of the effective atomic resonant
frequency $\protect\omega _{0}$ for the different dimensionless parameters $%
\protect\chi $. The initial state and the atomic number are chosen as $%
\left\vert J_{z}=-j\right\rangle $ and $N=100$, respectively.}
\label{fig5}
\end{figure}

From the above discussions, we argue that when increasing the dimensionless
parameter $\chi $ (from $\chi =-1$) or introducing the effective atomic
resonant frequency $\omega _{0}$ in the standard two-axis twisting
Hamiltonian $H_{\text{TAT}}$, the maximal squeezing factor $\xi _{\text{M}%
}^{2}$ increases, i.e., spin squeezing is reduced. Surprisingly, when we
control these two parameters simultaneously, spin squeezing can be enhanced
largely. To see this clearly, in Fig.~\ref{fig5} we numerically plot the
maximal squeezing factors $\xi _{\text{M}}^{2}$ of the generalized two-axis
spin Hamiltonian (\ref{TS}), i.e., $H=q(J_{x}^{2}+\chi J_{y}^{2})+\omega
_{0}J_{z}$, as a function of the effective atomic resonant frequency $\omega
_{0}$ for the different dimensionless parameters $\chi $, when the initial
state is chosen as $\left\vert J_{z}=-j\right\rangle $. This figure shows
that in the case of $\chi =-0.05$ or $\chi =-0.5$, when increasing the
effective atomic resonant frequency $\omega _{0}$, the maximal squeezing
factor $\xi _{\text{M}}^{2}$ first decreases largely and then increases,
i.e., spin squeezing is first enhanced largely and then reduced. For the
generalized one-axis twisting model $H_{\text{GOAT}}=qJ_{x}^{2}+\omega
_{0}J_{z}$, the maximal squeezing factor $\xi _{\text{M}}^{2}$ has a similar
behavior (see green dash-dot line in Fig.~\ref{fig5}) \cite{CKL01}, but its
magnitude cannot arrive at the order of our considered two-axis spin
Hamiltonian (\ref{TS}) because these two Hamiltonians have different
scalings with respect to the atomic number $N$. These results are benefit
for achieving the required spin squeezing in experiments.

\section{Discussions}

Finally, we estimate the relative parameters, based on recent experiments.
Two stable ground states are chosen respectively as $\left\vert
G_{0}\right\rangle =\left\vert F=1,m_{F}=1\right\rangle $ and $\left\vert
G_{1}\right\rangle =\left\vert F=2,m_{F}=2\right\rangle $ of ultracold
rubidium 87 atoms, where $F$ is the total angular momentum and $m_{F}$ is
the magnetic quantum number. It means that the atomic decay rate is $\gamma
/2\pi =3.0$ MHz, which is the same order as the photonic decay rate $\kappa
/2\pi =1.3$ MHz \cite{KB10,MPB14}. In such system, the atom-photon coupling
strengths can reach $g_{s_{1}}/2\pi =20$ MHz and $g_{s_{2}}/2\pi =15$ MHz,
respectively. When $\Omega _{s_{1}}/2\pi =20$ MHz and $\Delta _{s_{1}}/2\pi
=100$ MHz, which is responsible for deriving Eq.~(\ref{IP}), $\lambda
_{1}/2\pi =2$ MHz, and thus $q/2\pi =0.2$ MHz for $\omega _{A}/2\pi =-20$
MHz. This choice of the detuning $\omega _{A}$ also satisfies the dispersive
condition, which is a key condition to realize our generalized two-axis spin
Hamiltonian (\ref{TS}). For the dimensionless parameter $\chi $, it can be
easily controlled by tuning both the detuning $\omega _{B}$ and the Rabi
frequency $\Omega _{s_{2}}$. When $N=100$, numerical result shows that the
shortest time for generating the maximal squeezing factor $\xi _{\text{M}%
}^{2}$ is about $t_{m}\simeq 20$ ns [see, for example, in the insert part of
Fig.~\ref{fig4}], which is shorter than both the atomic and photonic
lifetimes $\gamma ^{-1}$ and $\kappa ^{-1}$. With the increasing of the
atomic number $N$ and the atom-photon coupling strength $g_{s_{1}}$, $t_{m}$
becomes shorter and shorter. This indicates that our proposal can be
accessible in the current experimental setups.

\section{Conclusions}

In summary, we have proposed an experimentally-feasible system, in which an
ensemble of ultracold six-level atoms interacts with two quantized cavity
fields and two pairs of Raman lasers, to realize a generalized two-axis spin
Hamiltonian $H=q(J_{x}^{2}+\chi J_{y}^{2})+\omega _{0}J_{z}$. We have
numerically calculated the experimentally-measurable maximal squeezing
factor and revealed that when $\omega _{0}=0$ and $\chi =-1$, the maximal
squeezing factor $\xi _{\text{M}}^{2}$ scales as $N^{-1}$. More importantly,
we have found that by combined with the dimensionless parameter $\chi (>-1)$%
, the effective atomic resonant frequency $\omega _{0}$ can enhance spin
squeezing largely. Our results are benefit for achieving the required spin
squeezing in experiments, and have a potential application in quantum
information and quantum metrology.

\section{Acknowledgements}

This work is supported in part by the 973 program under Grant
No.~2012CB921603; the NNSFC under Grant No.~11422433, No.~11434007,
No.~11447028, No.~61227902, and No.~61275211; the PCSIRT under Grant
No.~IRT13076; the NCET under Grant No.~13-0882; the FANEDD under Grant
No.~201316; ZJNSF under Grant No.~LY13A040001; OYTPSP; and SSCC.

\end{document}